\documentclass[12pt]{iopart}

\usepackage[utf8]{inputenc}
\usepackage{graphicx}
\usepackage{multicol}
\usepackage{wrapfig,lipsum,booktabs}
\usepackage{multirow}
\usepackage{xcolor}
\usepackage{colortbl,hhline}
\usepackage{adjustbox}
\usepackage{paralist}
\usepackage{amssymb}
\usepackage{makecell, multirow}

\begin{document}

\title[Modeling interdisciplinary interactions among PHY, MA and CS]{Modeling interdisciplinary interactions among Physics, Mathematics \& Computer Science}

\author{Rima Hazra\textsuperscript{*}, Mayank Singh\textsuperscript{+}, Pawan Goyal\textsuperscript{*}, Bibhas Adhikari\textsuperscript{*}, Animesh Mukherjee\textsuperscript{*}}
\address{\textsuperscript{*}Indian Institute of Technology Kharagpur, India, \textsuperscript{+}Indian Institute of Technology Gandhinagar, India }
\ead{to\_rima@iitkgp.ac.in}
\begin{abstract}
Interdisciplinarity has over the recent years have gained tremendous importance and has become one of the key ways of doing cutting edge research\footnote{\label{inter}https://physicstoday.scitation.org/doi/10.1063/PT.3.4164}. In this paper we attempt to model the citation flow across three different fields -- Physics (PHY), Mathematics (MA) and Computer Science (CS). For instance, is there a specific pattern in which these fields cite one another? We carry out experiments on a dataset comprising more than 1.2 million articles taken from these three fields. We quantify the citation interactions among these three fields through \textit{temporal bucket  signatures}. We present numerical models based on variants of the recently proposed \textit{relay-linking} framework to explain the citation dynamics across the three disciplines. These models make a modest attempt to unfold the underlying principles of how citation links could have been formed across the three fields over time.
\end{abstract}
\section{Introduction}
\label{sec:intro}
Since the last decade, research has been extensively performed by drawing ideas from various disciplines. Such research not only uncovers new directions which combine multiple existing disciplines but also transform them\footref{inter}. Applying concepts from one discipline to address problems of another discipline is becoming increasingly popular. Typically one observes such research in three different paradigms -- (i) multidisciplinary, (ii) interdisciplinary and (iii) transdisciplinary. In multidisciplinary research, researchers of different fields collaborate but they are limited to their research disciplines and use their domain knowledge to address the problem. In interdisciplinary research, researchers combine concepts of various disciplines to solve problems of their discipline. In case of transdisciplinary research, researchers create a holistic approach which includes research strategies beyond many disciplinary boundaries. In this work we focus on interdisciplinary research. The questions that we put word are -- how can one model the interdisciplinary interactions among different fields, do these interactions portray certain interesting patterns and could there be a way to explain the emergence of such patterns.
In this work, we model interdisciplinary research through citation exchange among various disciplines. In our experiments, we consider two basic science fields Mathematics (MA) and Physics (PHY) and one fast growing field Computer Science (CS). This article is an extended version of~\cite{Hazra:2019}. Note that in the rest of this paper we shall be using the terms ``discipline'' and ``field'' interchangeably.\\

Our objectives are the following.\\

\begin{enumerate}
    \item We use temporal bucket signatures to identify citation interaction patterns among the three fields of Physics, Mathematics and Computer Science.
    \item We study an array of baseline phenomenological model like simple preferential attachment and copying.
    \item Finally, we develop a phenomenological model based on the recently proposed \textit{relay linking} idea that quite accurately explains the emergence of the observed citation interactions.
\end{enumerate}

In~\cite{Hazra:2019} we have presented a first level empirical analysis of the interdisciplinary interactions across these three fields. Also, we observe how the citation flow changes from field to subfield level. We draw a few interesting observations. In case of Computer Sciences citing subfields of Physics, {\em Quantum Physics} remains at the top for a large time span. In early 1980s, quantum computing research had started its journey by holding the hands of physicist Paul Benio. Since last three decades, researchers had not only combined ideas of quantum mechanics but also {\em Machine learning}, {\em Artificial Intelligence}~\cite{Alvarez:2018} to build novel quantum
technologies\footnote{http://scitechconnect.elsevier.com/quantum-mechanics-change-computing/}\textsuperscript{,}\footnote{https://theconversation.com/how-quantum-mechanics-can-change-computing-80995}\textsuperscript{,}\footnote{https://www.newscientist.com/article/mg20827903-200-quantum-links-let-computers-understand-language/}.
In the context of citation flow from Mathematics to the subfields of Computer Science, in early years {\em Discrete Mathematics} was only cited. However, in recent years, {\em Information Theory} became the most cited and {\em Neural \& Evolutionary Computing Mathematics} and {\em Artificial Intelligence} came at second and third positions, respectively. We observed that during initial time-period, Physics highly cites
{\em Information Theory} papers of Computer Science. But over the time, it started citing
{\em Computational Complexity}, {\em Learning} and {\em Social and Information Networks} respectively. We observed that the Physics subfield {\em Physics and Society} (e.g.,~\cite{Gallotti:2016}) mostly interacts with {\em Social and Information Networks} (e.g.,~\cite{Liu:2014}). 
The most interesting observation was the growing interest of Physics (e.g., ~\cite{caravelli:2015}) and Mathematics (e.g.,~\cite{jorgensen:2016}) in the Computer Science subfield {\em Learning} (e.g.,~\cite{Carbajal:2015,xu:2011}). The major reason could be mathematicians and physicists have started research on formulating the mathematical foundations of various {\em machine learning} techniques. Physicists have been envisaging to solve complex problems in subfields like {\em Quantum Physics}, {\em Condensed Matter} with the help of {\em Learning}\footnote{https://qz.com/897033/applying-machine-learning-to-physics-could-be-\\the-way-to-build-the-rst-quantum-computer}. 

In this paper our primary objective is to develop phenomenological models to mimic the citation interactions sketched as temporal bucket signatures.

\subsection{Outline}
In section~\ref{sec:rel_weork}, we discuss the state-of-the-art studies in this area. In section~\ref{sec:dataset}, we describe the datasets in detail. In section~\ref{sec:models}, we attempt to explain the emergence of these patterns using various growth models. Section~\ref{sec:end} outlines the conclusion and future work.

\section{Related Work}
\label{sec:rel_weork}
Interdisciplinary nature of science is studied in different research fields like cognitive science~\cite{Till:2016,Kwon:2017}, social science~\cite{Pedersen:2016}, humanities~\cite{Pedersen:2016}, biology~\cite{Morillo:2003}, mathematics~\cite{Morillo:2003}, climate science~\cite{Weart:2012}, and several sub-fields of computer science~\cite{Chakraborty2018}. Many studies~\cite{Barthel:2017,Sayama:2012,Till:2016} have proposed quantifiable metrics to measure the degree of interdisciplinarity by utilizing factors such as scientific impact (h-index, i10-index, etc.), collaborator's knowledge (by measuring the entropy of topics), co-authors publication history, etc. In addition to research fields, the degree of interdisciplinarity is also measured at more granular levels such as individual articles, authors, and journals~\cite{Morillo:2003,Lariviere:2010}.

Bergmann {\em et al.}~\cite{Till:2016} propose a metric for quantifying interdisciplinarity of an article. The proposed metric is defined based on its authors’ research area (publications). They consider an article as `interdisciplinary' if the article's authors’ work on different research disciplines and have publications in different disciplines.   Lariviere {\em et al.}~\cite{Lariviere:2010} study how the interdisciplinary nature of an article helps in gathering rich scientific impact. They quantify the degree of interdisciplinarity of an article as the fraction of its cited references published in other discipline journals. Surprisingly, they observe that interdisciplinary articles have a low scientific impact. Whereas Yegros {\em et al.}~\cite{Yegros-Yegros:2015} analyze the effect of the degree of interdisciplinarity on the citation impact of individual articles in the field of cell biology (CBIOL), engineering -- electrical \& electronic (EEE), food science and technology (FSTA) and physics -- atomic, molecular \& chemical (Physics-AMC). Each article consists of multiple concepts, data, algorithms from different disciplines. They explore different aspects of diversity present in an article. They measure the diversity based on three categories, i.e., the number of distinct disciplines, evenness of distribution of disciplines and degree of difference among different disciplines. They observe that combining multiple disciplines, with a high degree of similarity, yields good citation impact. They observe that the scientific community reluctantly cite articles that encompass multiple disciplines with low similarity. Chen {\em et al.}~\cite{Chen:2015} analyze the degree of interdisciplinarity of highly cited articles. Their analysis shows that interdisciplinary research is more impactful than monodisciplinary (works in a single discipline) research.\\

Bonaventura {\em et al.}~\cite{Bonaventura:2017} conduct a comparison between the success of monodisciplinary and interdisciplinary researchers. Unlike Bergmann {\em et al.}~\cite{Till:2016}, they observe interdisciplinary researchers are more successful than the monodisciplinary researchers. The success of a researcher is measured by the heterogeneity of researchers' research topics and the diversity of topics which are gathered during collaboration. Barthel {\em et al.}\cite{Barthel:2017} explore the degree of interdisciplinary collaboration between natural and social sciences. They leveraged on a set of articles on groundwater research over the period 1990--2014 for this study. They investigate research topic diversity of authors, article title, references, etc. They observe that journals publishing interdisciplinary articles are relatively less impactful. Sayama {\em et al.}~\cite{Sayama:2012} constructed a network which shows connection among researchers as well as their research disciplines. They proposed a measure called {\it visibility boost} which defines the relatedness between researchers and their research disciplines. Two researchers are connected in the network if they have high visibility boost for the same research discipline. They observed that the diversity of research disciplines of a researcher is highly correlated with her/his centrality value in the network.\\

Leydesdorff {\em et al.}~\cite{Leydesdorff:2011} constructed a network of journals and studied network metrics (betweenness centrality), journal indicators (Shannon entropy and Gini coefficient based on journal’s citation distribution), and Rao-Stirling index for quantifying interdisciplinary nature. Their study shows that these metrics capture different aspects of interdisciplinarity. Soloman {\em et al.}~\cite{Solomon:2016} investigated the degree of multidisciplinarity of popular journals such as {\it Nature} and {\it Science}. They consider three disciplines cell biology, physical chemistry and cognitive science and find that articles published in these journals are highly interdisciplinary. They compared the articles published in monodisciplinary popular journals with articles published in {\it Nature} and {\it Science}. They find articles published in {\it Nature} and {\it Science} are more interdisciplinary than articles published in monodisciplinary popular journals.

To the best of our knowledge, none of the above works have presented the dynamics of citation interactions among these three different disciplines. Rather, the focus has been very specific to measuring interdisciplinarity debarring  consideration of evolution of the concept of interdisciplinarity. \\

In our earlier paper~\cite{Hazra:2019}, our main focus was to observe the patterns of citation flow from one field to another field and this flow changes over timescale. In addition, we observed how some of the popular subfields of these fields  became less popular and new subfields emerged over time. 

Here we advance this work to develop an array of phenomenological models to mimic the empirical observations in the form of temporal bucket signatures we made in~\cite{Hazra:2019}. 

\section{Datasets}
\label{sec:dataset}
We have automatically collected research articles from \textit{arXiv}\footnote{www.arxiv.org}, one of popular pre-print repositories. Our dataset includes 1.2 million articles published in nine major fields and submitted between 1990--2017. The nine fields are Physics, Mathematics, Computer Science, Quantitative Biology, Quantitative Finance, Statistics, Electrical Engineering and Systems Science, and Economics. The number of papers from each of these fields is noted in Table~\ref{tab:dataset_allfield}.
\begin{table}[h]
\centering
\caption{Description of the arXiv data.}\label{tab:dataset_allfield}

\begin{tabular}{c c c c} \hline
{\bf Sl. No} & {\bf Fields} & {\bf \# papers} & {\bf \# subfields} \\ \hline
1 & Computer Science & 1,41,662 & 40\\ \hline
2 & Physics & 8,04,360 & 50 \\ \hline
3 & Mathematics & 2,84,540 & 33 \\ \hline
4 & Quantitative Biology & 26708 & 10 \\ \hline
5 & Quantitative Finance & 1878 & 9 \\ \hline
6 & Statistics & 43100 & 6\\ \hline
7 & Electrical Engineering & 239 & 4 \\
& and System Science & & \\ \hline
8 & Economics & 35 & 3\\ \hline
\end{tabular}
\end{table}
There is a list of subfields available for a given field on arxiv~\footnote{https://arxiv.org/}. In the field of physics, for instance, the subfields are {\em Computational Physics} (physics.comp-ph), {\em Instrumentation and Methods for Astrophysics} (astro-ph.IM). We further map the fields into different subfields. Also, for each paper, the respective subfields are present in the arxiv metadata. 

For our experiments we shortlist the following three fields based on the number of largest number of data points available -- Computer Science ($CS$), Mathematics ($MA$) and Physics ($PHY$) (see Table~\ref{tab:dataset_allfield}). We construct the citation network using the references present in the ``.bbl'' files. For every article we only consider those referenced articles that are present within \textit{arXiv} through a simple string matching of article titles. The major challenge we face is finding the article in the arxiv. In a few cases, the articles are directly published after not being uploaded to arXiv. This problem is especially evident in the fields of Mathematics ({\em MA}) and Physics ({\em PHY}), where publishing publications on arXiv for reproducibility is less common than in Computer Science ({\em CS}). In our current experiment, we ignore these special cases.

In the citation graph, we have in total 322028 number of nodes (number is less because of multiple steps of reference mapping). Total number of citation edges in the graph is 256838. We discard articles that have less than five extracted references. 

\noindent\textbf{Approximating citation interaction}: We carry out in-depth analysis of citation interactions among two basic science fields and one fast growing field. Citations received from papers of the same field are defined as \textit{self-field citations}. Citation gained from papers of other fields are defined as \textit{non self-field citations}. The proportion of self-field and non self-field citations are noted in Table~\ref{tab:self_nonSelf}.

\begin{table}[h]
\centering
\caption{Proportion of self and non self-field citations flows from these three fields.}\label{tab:self_nonSelf}
\begin{tabular}{c c c c} \hline
{\bf Sl. No} & {\bf Fields} & {\bf \% Self-field} & {\bf \% Non self-field} \\ 
 &  & {\bf citations} & {\bf citation} \\ \hline
1 & Computer Science & 84.08\% & 15.917\%\\ \hline
2 & Mathematics & 89.50\% & 10.49\%\\ \hline
3 & Physics & 97.12\% & 2.872\% \\ \hline
\end{tabular}
\end{table}

\section{Modeling citation flow} 
\label{sec:models}

\subsection{Temporal bucket signatures}
Temporal bucket signatures (TBS)~\cite{Singh:2017} are a new way of visualising the citation interactions that take place over time. These are stacked histograms representing the `age' of the target paper cited in some source paper. Let us assume that the citation links are organised into time buckets of width $T$ (say $T=5$ years). One can always partition the entire set of papers in the dataset into these buckets based on their year of publication. For a given bucket, we can compute the fraction of citations going from that bucket to the previous buckets. For any individual bucket, these fraction of citations received from the following buckets are stacked over one another to represent the temporal bucket signature. Note that self-field citations are way higher in number than non self-field citations so we prepare separate TBSs for the former and latter categories. To be more precise, for the former case, we study how the papers in the older buckets of the same field receive citations from the younger bucket papers of the same field. On the other hand, in the latter case we study how the papers in the older buckets of a particular field receive citations from the younger bucket papers of another field.

\begin{figure*}[tbhp]
\centering
\includegraphics[width = \textwidth]{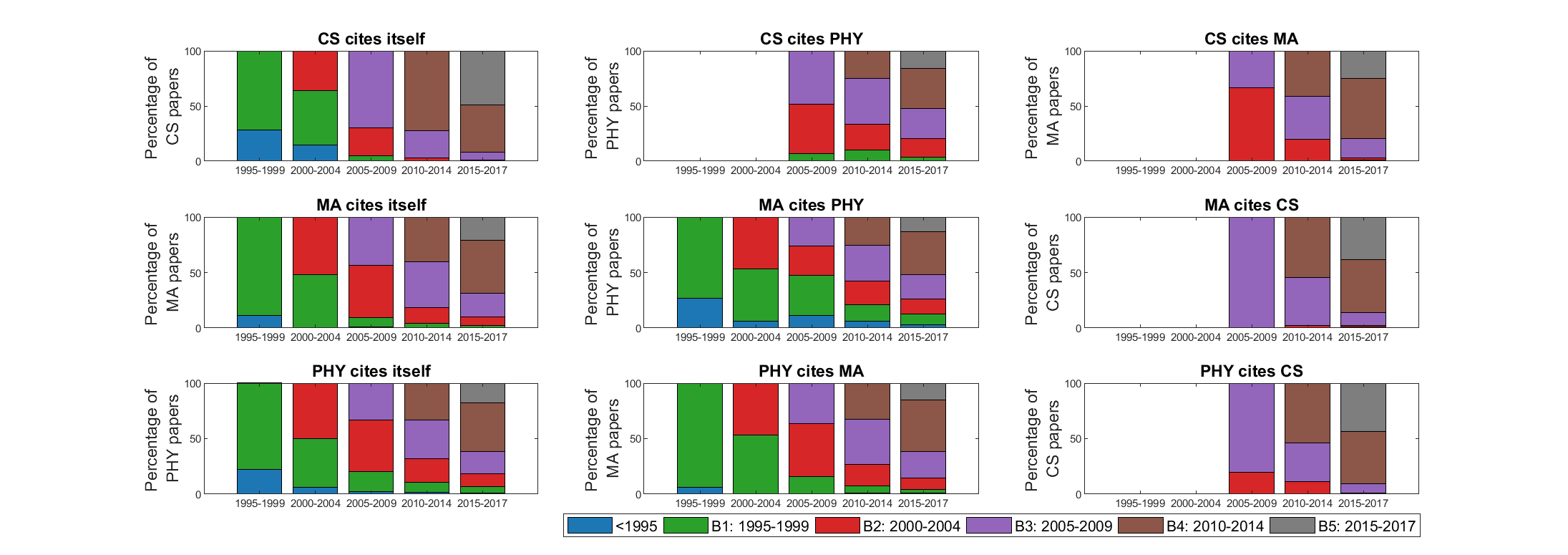}
\caption{Temporal bucket signatures for self- and non self-field citations.}
\label{fig:3field_temporal}
\end{figure*}

\noindent\textbf{Key observations}: In Figure~\ref{fig:3field_temporal} we present the TBSs. An intriguing observation is that while $CS$ papers typically cite recent papers from their own field, they tend to cite older papers of the other two fields. On the other hand, papers from both $MA$ and $PHY$ tend to cite older papers of their own field and newer papers from $CS$. This possibly is a manifestation of these fields function and grow over time.


\subsection{Phenomenological models}
Motivated by the inter-field citation interactions, we propose to model this interesting behavior. We found vast literature on growth models of citation networks, ranging from standard preferential attachment model~\cite{albert2002statistical} to recently proposed relay-link based models~\cite{Singh:2017}. However, to the best of our knowledge, there is no work on inter-field citation growth modeling. We, therefore, adapt the previous models to fit this problem and compare them based on temporal bucket signatures. 

The modeling experiment is conducted in two phases. In the first phase, we create a warm-up dataset to avoid cold-start problem during model initialization. In our experiments, we consider warm-up dataset that consists of citation information generated between 1995--2009. The second phase involves model simulation between the year range 2010--2017. We, verbatim, follow the simulation guidelines provided by~\cite{Singh:2017}. In addition, we also keep track of the field information of each paper entering into the system. Empirically, we found that average in-degree of each paper is eight. Therefore, we model each incoming paper entering the system with eight outgoing edges. For each incoming paper, we sample its field by utilizing the field distribution from the empirical data. For each outgoing edge of an incoming paper, we select the destination node by simulating following growth models.

\subsubsection{Preferential attachment (PA)}  
\begin{figure*}[t]
\centering
\includegraphics[width = \textwidth]{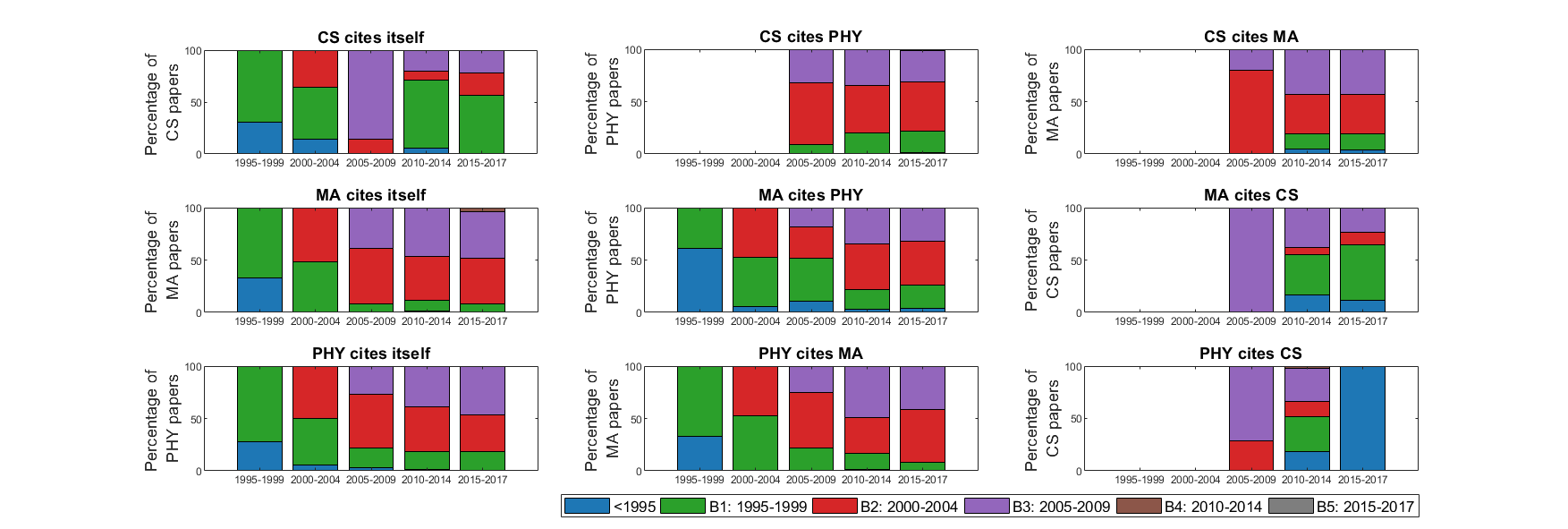}
\caption{\textbf{PA}: Temporal bucket signatures generated using \textit{PA}. The results do not seem to mimic the real data (Figure \ref{fig:3field_temporal}) well.}
\label{fig:PA_TYPE1_temporal}
\end{figure*}

Albert and Barabasi's~\cite{albert2002statistical} seminal proposal of a mechanistic model to capture ``rich gets richer'' phenomenon laid the foundation stone of probabilistic growth models. A high degree node has more probability to attract new incoming links than low degree nodes. When an incoming node enters the system along with a set of outgoing edges, then for each edge, it first samples destination field from the empirical field distribution. Next, it preferentially samples (based on in-degree) the destination node among the candidate papers published in the previously sampled destination field. The temporal bucket signatures for this model is shown in Figure~\ref{fig:PA_TYPE1_temporal}.

\subsubsection{Copying models (CP)}
In these models~\cite{Kumar:2000}, similar to $PA$, we first sample destination field from the empirical field distribution and then preferentially sample a node from the chosen field (\textit{sampling phase}). In addition, we copy references of the destination paper (\textit{copying phase}). The sampling and copying phase continue iterating alternatively until all outgoing edges of an incoming paper have received their respective destination nodes. We experiment with three different variations of copying models that differ in the copying phase. 

\begin{compactitem}
\item {\bf In field copying model ($ICP$):} If the sampled destination field (after sampling phase) matches with the field of the incoming node, it copies all references (having the same field) of the destination paper too (\textit{copying phase}).  If the field of the destination node does not match with the field of the incoming node, no copying takes place. The temporal bucket signatures for this model is shown in Figure~\ref{fig:PA_TYPE2_temporal}.

\begin{figure*}[h]
\centering
\includegraphics[width = \textwidth]{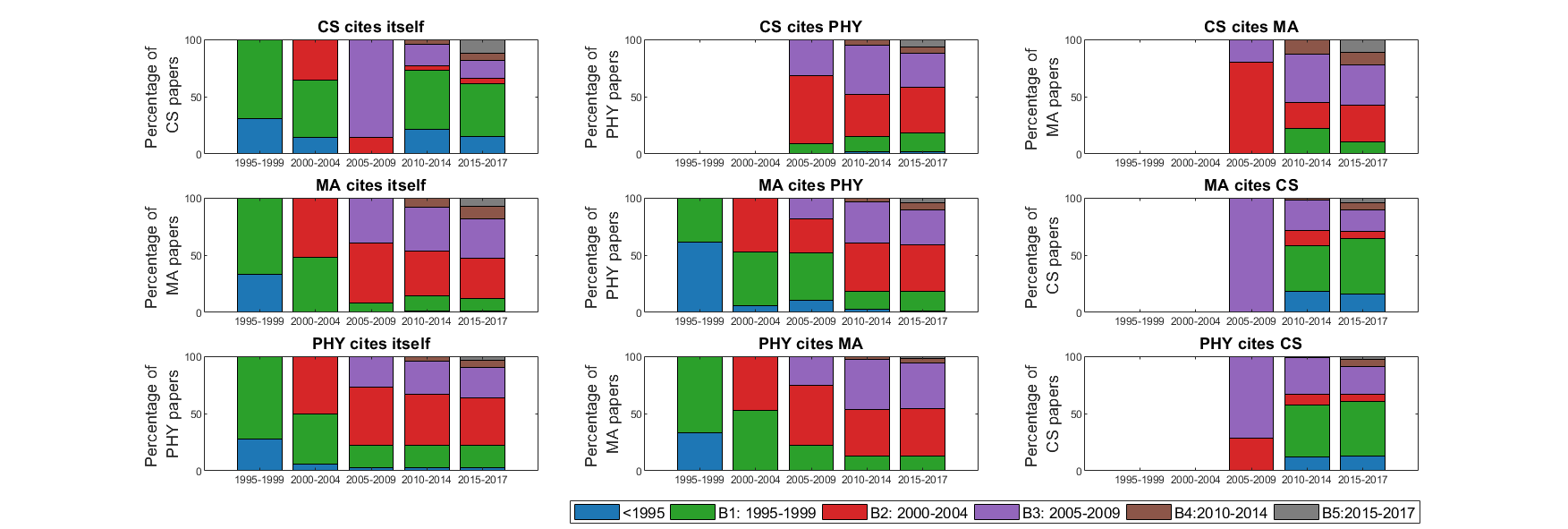}
\caption{\textbf{ICP:} Temporal bucket signatures generated using \textit{ICP}. Once again, the results do not seem to mimic the real data (Figure \ref{fig:3field_temporal}) well.}
\label{fig:PA_TYPE2_temporal}
\end{figure*}

\item \noindent{\bf All field copying model ($ACP$):} This model works quite similar to the previous one, with an additional feature that during copying phase all references are copied irrespective of their field. The temporal bucket signatures for this model is shown in Figure~\ref{fig:PA_TYPE3_temporal}.

\begin{figure*}[h]
\centering
\includegraphics[width = \textwidth]{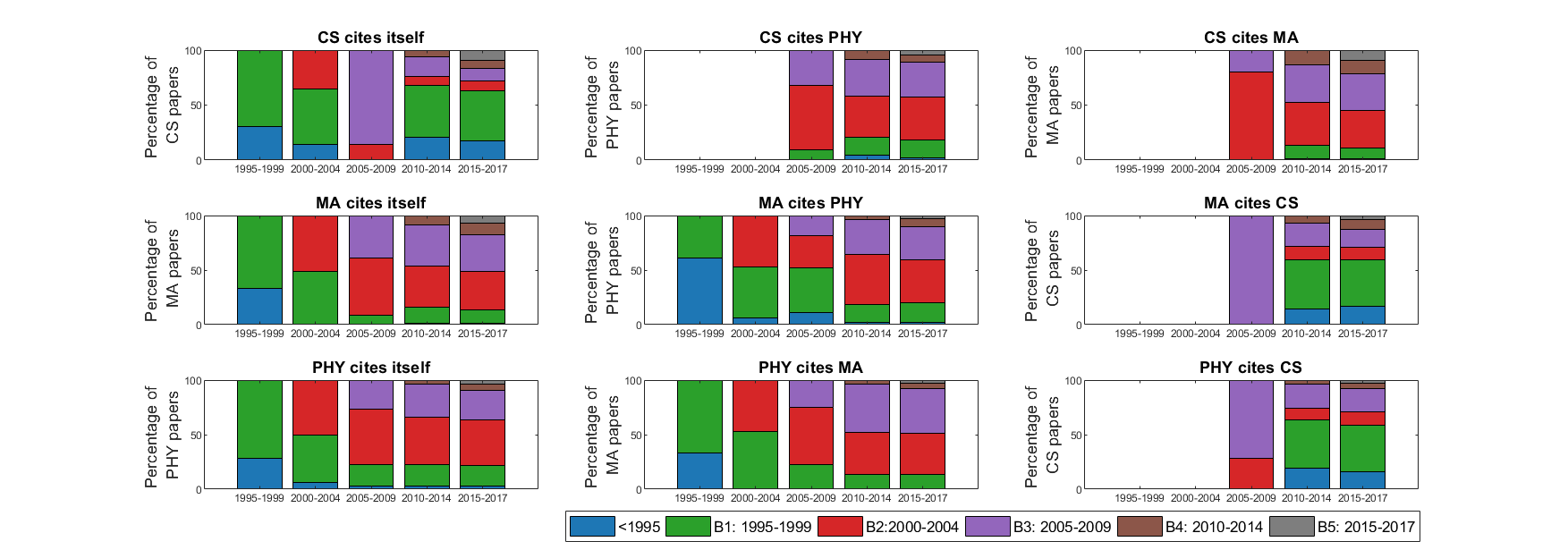}
\caption{\textbf{ACP:} Temporal bucket signatures generated using \textit{ACP}. The trends seem to start getting similar to the ground truth (Figure \ref{fig:3field_temporal}).}
\label{fig:PA_TYPE3_temporal}
\end{figure*}
\item \noindent{\bf Random all field copying model ($RACP$):} In this model, we use both the infield and the outfield reference selection procedure as in the $ACP$ approach. However, we choose the outfield papers from the bucket of outfield papers cited by the incoming node's field. If there is no such outfield paper that the incoming node's field is citing, then we preferentially choose one of the top cited papers from that outfield as the destination node. The temporal bucket signatures for this model is shown in Figure~\ref{fig:PA_TYPE4_temporal}.

\begin{figure*}[h]
\centering
\includegraphics[width = \textwidth]{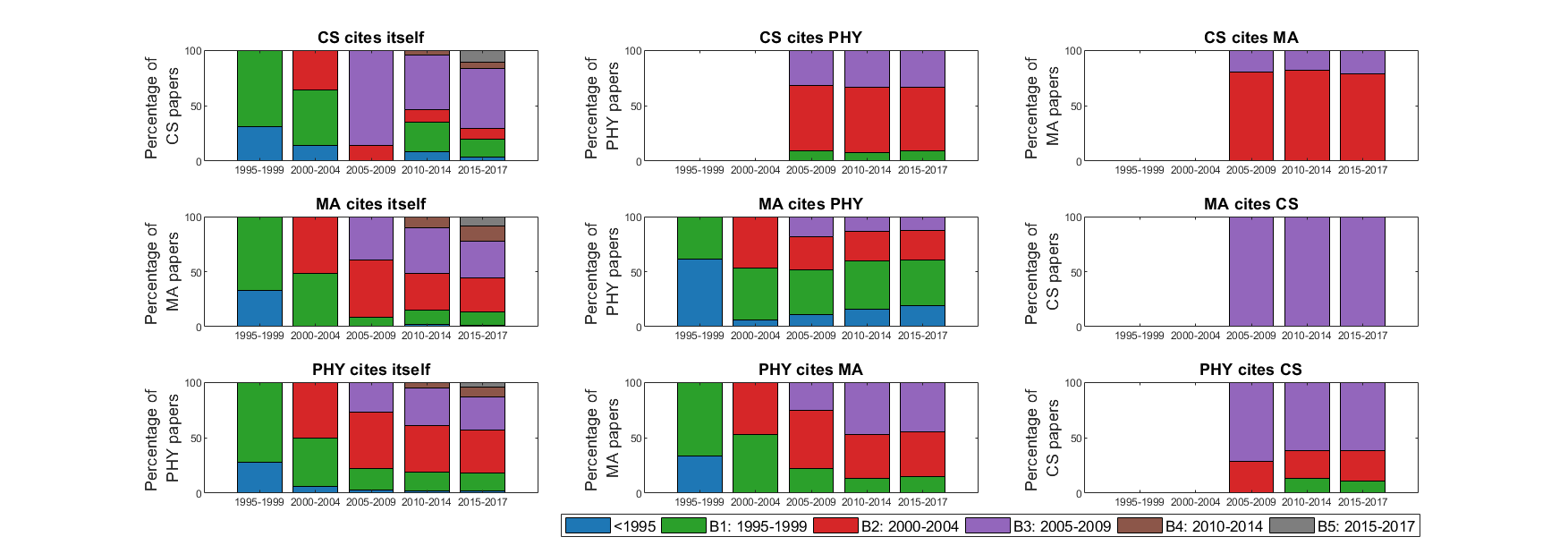}
\caption{\textbf{RACP:} Temporal bucket signatures generated using \textit{RACP}. The trends are again quite similar to the ground truth (Figure \ref{fig:3field_temporal}).}
\label{fig:PA_TYPE4_temporal}
\end{figure*}
\end{compactitem}

\begin{table}[ht]
\centering
\caption{L1 Distance for all the models. Note that we set the $\theta$ and the $\lambda$ for the in field citations to the same values as in~\cite{Singh:2017}; we choose $\theta=1$ and $\lambda=1$ for the out field citations since this results in the least L1 distance.}\label{tab:L1Distance}
\tiny
\begin{tabular}{*{9}{c}} \hline
\multirow{2}{*}{\centering \textbf{Citation}} & \multicolumn{8}{c}{\bf Models}\\ 
\cline{2-9}
 & {\em \bf PA} & {\em \bf ICP} &{\em \bf ACP} & {\em \bf RACP} & {\em \bf IIPRC} & {\em \bf OIPRC} & {\em \bf BIPRC} & {\em \bf CIPRC}\\ \hline
 $CS\rightarrow MA$ & 2.65 & 2.08 & 2.08 &    3.08 &    3.04 & 3.59& \cellcolor{yellow!20}1.98& \cellcolor{green!20}1.94\\ \hline
 $CS\rightarrow PHY$ & 1.97 & 1.68 & 1.69 & 2.11 & 2.25 & 1.89 &\cellcolor{yellow!20}1.64& \cellcolor{green!20}1.52\\ \hline
 $MA\rightarrow CS$ & 3.20 & 2.87 & 2.89 & 2.90 & 2.71 & 2.90 &\cellcolor{green!20}2.59& \cellcolor{yellow!20}2.65\\ \hline
 $MA\rightarrow PHY$ &2.48 & 2.17 & 2.17 & 2.95 &2.90 & 2.75 & \cellcolor{yellow!20}2.22&\cellcolor{green!20} 2.09\\ \hline
 $PHY\rightarrow CS$  & 3.11 & 3.03 & 3.03 & 3.06 & 3.06 & 3.06 & \cellcolor{yellow!20}2.65& \cellcolor{green!20}2.58\\ \hline
 $PHY\rightarrow MA$  & 2.67 & 2.47 & 2.42 & 2.64 & 2.61 & 2.66 &\cellcolor{green!20}2.38& \cellcolor{green!20}2.38\\ \hline
 $CS\rightarrow$ self  &3.72 & 3.32 & 3.33 & \cellcolor{pink}3.16 & \cellcolor{pink}3.16 & 3.24 &\cellcolor{yellow!20}2.94& \cellcolor{green!20}2.90\\ \hline
 $MA\rightarrow$ self  & 2.70 & 2.25 & 2.26 & \cellcolor{pink}2.12 & \cellcolor{pink}2.01 & 2.13  &\cellcolor{green!20}2.06& \cellcolor{green!20}2.06\\ \hline
 $PHY\rightarrow$ self & 2.17 & 1.96 & 1.96 & \cellcolor{pink}1.80 & \cellcolor{pink}1.79 & 1.80 &\cellcolor{yellow!20}1.79& \cellcolor{green!20}1.78\\ \hline
 Overall& 2.28 & 2.025 & 2.02 & 1.89 & 1.89 & 1.91 & \cellcolor{yellow!20}\textbf{1.85} & \cellcolor{green!20}\textbf{1.84} \\ \hline
\end{tabular}
\end{table}

\subsubsection{Relay-link models}
The prime motivation behind the relay-link model~\cite{Singh:2017} proposal is to better understand the obsolesce phase of research papers. The authors in their work show, that the incoming links to an older paper are diverted to its younger child (a paper which cites the older paper). In brief, they proposed a hypothesis that when an incoming node enters, it samples a destination paper preferentially. If the sampled paper is too old (modeled by exponential distribution parameterized by $\lambda$), the model proceeds for a `relay' controlled by another parameter $\theta$. A higher $\lambda$ leads to early aging and a higher $\theta$ leads to high chances of relaying. If successful, then the link is relayed to its child paper. Link-relaying can happen recursively. However, the original proposal does not deal with inter-field interactions. We, therefore, present several adaptations of the relay-linking variant -- \textit{iterated preferential relay cite} ($IPRC$) to model inter-field interactions. Here, we follow the process of tuning the parameters $\theta$ and $\lambda$ from paper~\cite{Singh:2017}.

\begin{compactitem}
\item \noindent{\bf In field IPRC ($IIPRC$):}  In this relay-link variation, we perform relay only for infield destination nodes. To select outfield destination nodes, we adopt a two-step process. In the first step, we preferentially select outfield papers from the already selected in field papers' outfield references. In case the first step fails to find any relevant paper, we proceed to the second step. Here, we preferentially select papers from all the outfield references of the incoming paper's field. The temporal bucket signatures for this model is shown in Figure~\ref{fig:RELAY_TYPE5_temporal}.

\begin{figure*}[h]
\centering
\includegraphics[width = \textwidth]{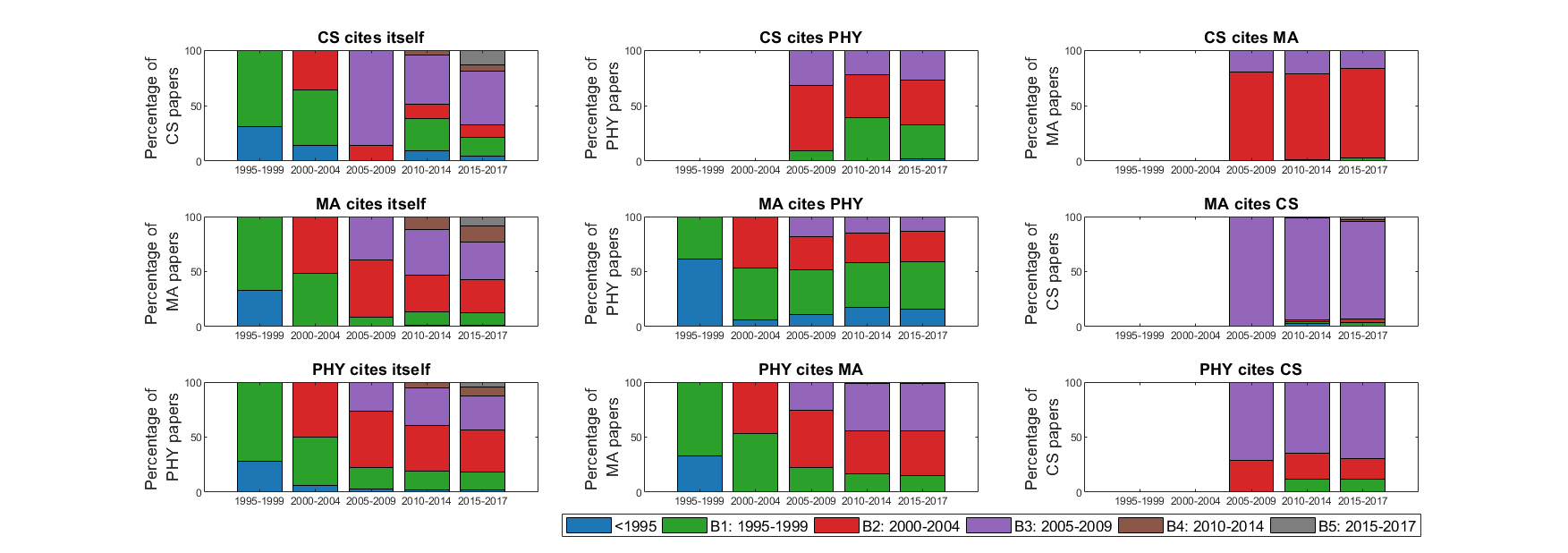}
\caption{\textbf{IIPRC:} Temporal bucket signatures generated using \textit{IIPRC}. The trends start getting quite close to the ground truth (Figure \ref{fig:3field_temporal}).}
\label{fig:RELAY_TYPE5_temporal}
\end{figure*}


\item \noindent{\bf Outfield IPRC ($OIPRC$):} In this model, we perform relay for both in-field and out-field destination nodes. However, the initial outfield destination node selection process remains the same as the previous version ($IIPRC$). We keep the $\theta$ and $\lambda$ values for the infield citations same as in~\cite{Singh:2017}; we choose $\theta=1$ and $\lambda=1$ for the outfield citations since this results in the least L1 distance. The temporal bucket signatures for this model is shown in Figure~\ref{fig:RELAY_TYPE6_temporal}.

\begin{figure*}[h]
\centering
\includegraphics[width = \textwidth]{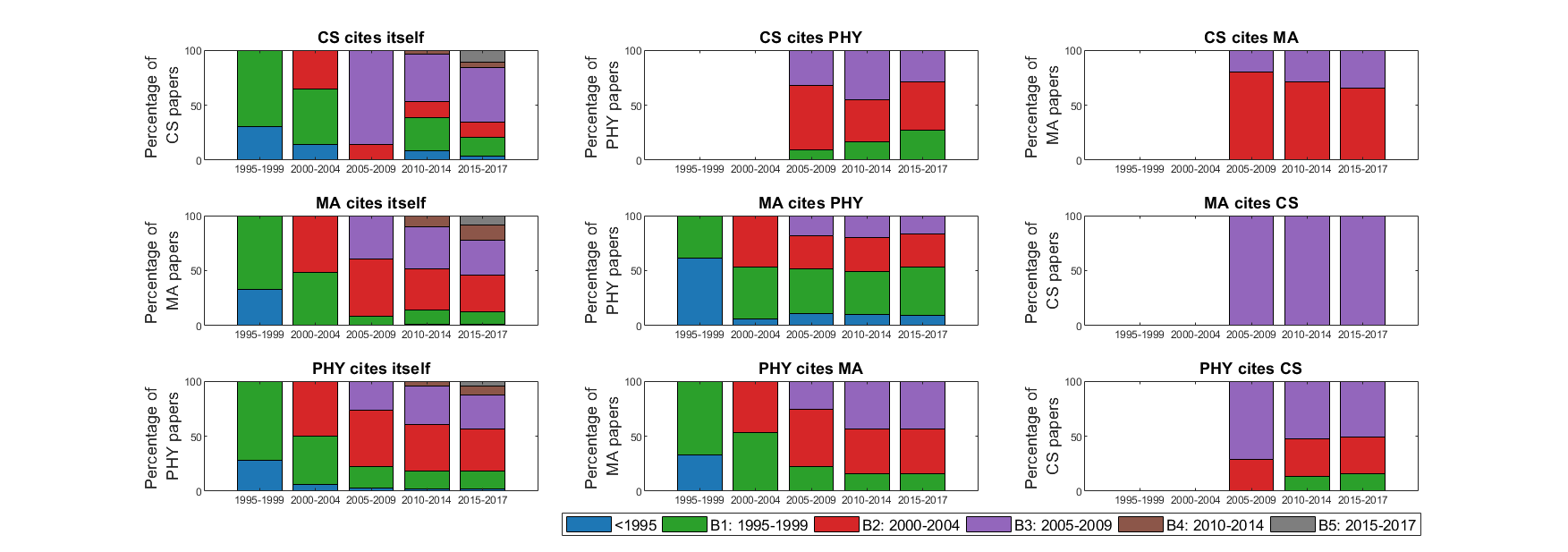}
\caption{\textbf{OIPRC:} Temporal bucket signatures generated using \textit{OIPRC}. The trend again seems to be quite close to the ground truth (Figure \ref{fig:3field_temporal}).}
\label{fig:RELAY_TYPE6_temporal}
\end{figure*}

\item \noindent{\bf Both field IPRC ($BIPRC$):} In this model, we again perform relay for both infield and outfield destination nodes. In case of infield as well as outfield destination nodes, we preferentially select papers and relay to infield and outfield papers respectively. The temporal bucket signatures for this model is shown in Figure~\ref{fig:RELAY_TYPE7_temporal}. 

\begin{figure*}[h]
\centering
\includegraphics[width = \textwidth]{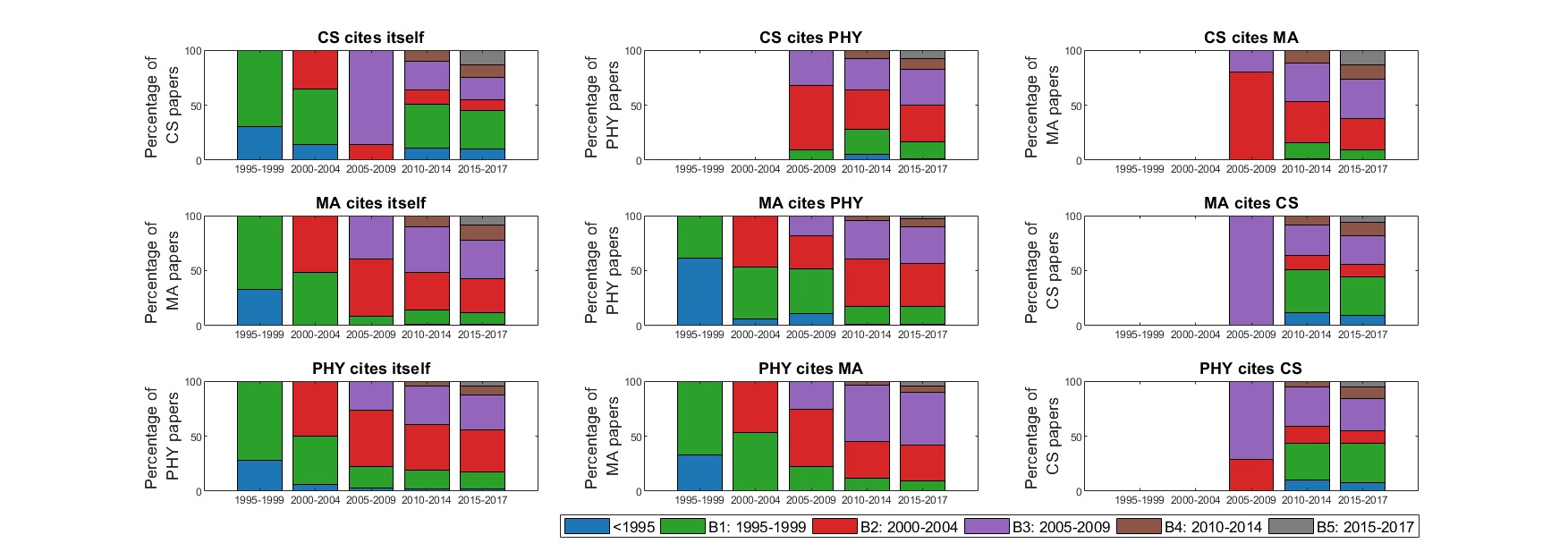}
\caption{\textbf{BIPRC:} Temporal bucket signatures generated using \textit{BIPRC}. The trend again seems to be quite close to the ground truth (Figure \ref{fig:3field_temporal}).}
\label{fig:RELAY_TYPE7_temporal}
\end{figure*}

\item \noindent{\bf Copy IPRC ($CIPRC$):} In this model, we merge $ICP$ and $IIPRC$, i.e., we select in field destination nodes by leveraging $IIPRC$. The outfield destination nodes are selected by $ICP$'s copying mechanism. Temporal bucket signature for this model is given in Figure~\ref{fig:RELAY_TYPE8_temporal}.

\begin{figure*}[h]
\centering
\includegraphics[width =\textwidth]{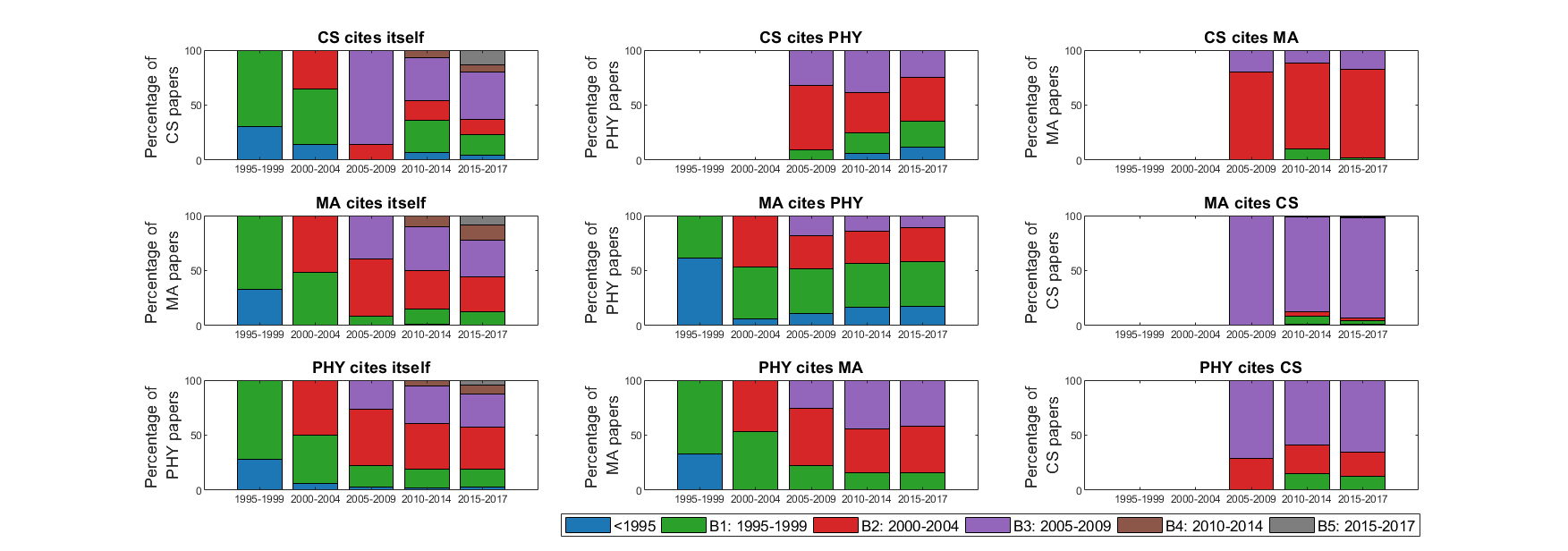}
\caption{\textbf{CIPRC:} Temporal bucket signatures generated using \textit{CIPRC}. Visually, the stacked plots here seem to be the closest to the ground truth (Figure \ref{fig:3field_temporal}).}
\label{fig:RELAY_TYPE8_temporal}
\end{figure*}
\end{compactitem}

\subsection{Observations}
We, next discuss the key results. We report our findings in Table~\ref{tab:L1Distance} by comparing the temporal bucket signatures obtained from the proposed models and the real data using L1 distance. From Table~\ref{tab:L1Distance}, it is clear that the basic $PA$ model is not able to reproduce the temporal bucket signatures faithfully. Both $ICP$ and $ACP$ are almost equidistant from ground truth. While $RACP$ models self field citations well it does not do well with the non self field citations. We observed that relay-link models (specially $BIPRC$ and $CIPRC$) are able to capture well the interaction among different field in ground truth. In paper~\cite{Singh:2017}, authors propose a model which takes into account the obsolescence of the research papers over time. Our intuition is that the relay-based models can model the citation interaction among different field closely to the ground truth.
$IIPRC$ show similar results as $RACP$. The models $OIPRC$, $BIPRC$ and $CIPRC$ gradually result in lower L1 distances in that order. The $CIPRC$ model, in particular, show reasonably acceptable performance for both the self-field and non self-field citations. In fact, this model achieves the lowest L1 distance from the ground truth.

\subsection{Application}

This work provides an empirical evidence of the rise and evolution of interdisciplinary research over the years, and examines applicability of existing network models to replicate the citation data. Moreover, the citation data reveals the emergence of subfields which inherit tools to be applied to other fields or can grow by borrowing ideas from other disciplines. This has immediate implications in identifying potential collaborators from other disciplines that can help developing interdisciplinary research projects.

Multiple studies~\cite{Bromham:2016,Rons:2013,Kwon:2017} confirm that interdisciplinary research receives lesser funding  than monodisciplinary research. The possibility of receiving funds is inversely proportional to the degree of interdisciplinarity involved in the research\footnote{https://www.the-scientist.com/daily-news/interdisciplinary-research-attracts-less-funding-33290}. Researchers assume two probable reasons. First, funding agencies are not well-aware of the recent interdisciplinary research, for example, how research in $PHY$ involves algorithms, concepts from subfields of $CS$. Second is the lack of expert peer reviewers for an interdisciplinary research proposal. We analyze how the basic fields ($PHY$ and $MA$) cite various subfields of $CS$ at different time duration. This analysis can potentially help funding agencies to identify recent popular interdisciplinary research. Our study can help funding agencies to assess funding applications by visualizing the impact of recent interdisciplinary research at different time-spans. Our work can help eliminate the organizational constraints that limit the capacity of the funding agencies to fully embrace novel ways of interdisciplinary collaboration and investigation and fund them. In an alternative perspective, the current study can help young researchers to choose the appropriate fields that would allow them to acquire larger visibility, citations as well as draft successful fund proposal~\cite{bridle:13}.  


\section{Conclusion and future work}
\label{sec:end}
In this work, we explore research articles published in between 1995-2017 in three different research disciplines -- \textit{PH}, \textit{MA}, and \textit{CS} and represent the citation interactions using temporal bucket signatures. We presented an array of simulation models based on variants of the recently proposed relay-linking mechanics to explain the citation dynamics across the three disciplines. In future, we plan to expand this study to other scientific fields. 

We admit that the models presented here are not treated analytically; however we also would like to point out the these model are extremely complex and the approximations that one needs to do make the analytical treatment possible can potentially render the model ineffective in capturing the real dynamics of citation interactions. We wish to investigate in future what realistic approximations are possible to allow for any analytical treatment thereof.

\section{References}
\bibliographystyle{plain}
\bibliography{sn-bibliography}

\begin{thebibliography}{10}

\bibitem{albert2002statistical}
R{\'e}ka Albert and Albert-L{\'a}szl{\'o} Barab{\'a}si.
\newblock Statistical mechanics of complex networks.
\newblock {\em Reviews of modern physics}, 74(1):47, 2002.

\bibitem{Alvarez:2018}
U.~Alvarez-Rodriguez, M.~Sanz, L.~Lamata, and E.~Solano.
\newblock Quantum artificial life in an ibm quantum computer.
\newblock 8, 2018.

\bibitem{Barthel:2017}
Roland Barthel and Roman Seidl.
\newblock Interdisciplinary collaboration between natural and social sciences
  – status and trends exemplified in groundwater research.
\newblock {\em PLOS ONE}, 12(1):1--27, 01 2017.

\bibitem{Till:2016}
Till Bergmann, Rick Dale, Negin Sattari, Evan Heit, and Harish~S Bhat.
\newblock The interdisciplinarity of collaborations in cognitive science.
\newblock {\em Cognitive science}, 41(5):1412--1418, 2017.

\bibitem{Bonaventura:2017}
M.~{Bonaventura}, V.~{Latora}, V.~{Nicosia}, and P.~{Panzarasa}.
\newblock The advantages of interdisciplinarity in modern science.
\newblock {\em ArXiv e-prints}, December 2017.

\bibitem{bridle:13}
Helen Bridle, Anton Vrieling, Monica Cardillo, Yoseph Araya, and Leonith
  Hinojosae.
\newblock Preparing for an interdisciplinary future: A perspective from
  early-career researchers.
\newblock {\em Future}, 53:22--32, 2013.

\bibitem{Bromham:2016}
Lindell Bromham and Russell Dinnage.
\newblock Interdisciplinary research has consistently lower funding success.
\newblock {\em Nature}, 534, 2016.

\bibitem{caravelli:2015}
Francesco Caravelli.
\newblock Trajectories entropy in dynamical graphs with memory, 2015.

\bibitem{Carbajal:2015}
Juan~Pablo Carbajal, Joni Dambre, Michiel Hermans, and Benjamin Schrauwen.
\newblock Memristor models for machine learning.
\newblock {\em Neural Computation}, 27(3):725–747, Mar 2015.

\bibitem{Chakraborty2018}
Tanmoy Chakraborty.
\newblock Role of interdisciplinarity in computer sciences: quantification,
  impact and life trajectory.
\newblock {\em Scientometrics}, 114(3):1011--1029, Mar 2018.

\bibitem{Gallotti:2016}
Riccardo Gallotti, Armando Bazzani, Sandro Rambaldi, and Marc Barthelemy.
\newblock A stochastic model of randomly accelerated walkers for human
  mobility.
\newblock {\em Nature Communications}, 7(1), Aug 2016.

\bibitem{Hazra:2019}
Rima Hazra, Mayank Singh, Pawan Goyal, Bibhas Adhikari, and Animesh Mukherjee.
\newblock The rise and rise of interdisciplinary research: Understanding the
  interaction dynamics of three major fields - physics, mathematics and
  computer science.
\newblock In {\em Digital Libraries at the Crossroads of Digital Information
  for the Future - 21st International Conference on Asia-Pacific Digital
  Libraries, {ICADL} 2019, Kuala Lumpur, Malaysia, November 4-7, 2019,
  Proceedings}, pages 71--77, 2019.

\bibitem{jorgensen:2016}
Palle Jorgensen and Feng Tian.
\newblock Nonuniform sampling, reproducing kernels, and the associated hilbert
  spaces, 2016.

\bibitem{Kumar:2000}
Ravi Kumar, Prabhakar Raghavan, Sridhar Rajagopalan, D~Sivakumar, Andrew
  Tomkins, and Eli Upfal.
\newblock Random graph models for the web graph.
\newblock {\em In FOCS}, page 57–65, 2000.

\bibitem{Kwon:2017}
Seokbeom Kwon, Gregg E.~A. Solomon, Jan Youtie, and Alan~L. Porter.
\newblock A measure of knowledge flow between specific fields: Implications of
  interdisciplinarity for impact and funding.
\newblock {\em PLOS ONE}, 12(10):1--16, 10 2017.

\bibitem{Leydesdorff:2011}
Loet Leydesdorff and Ismael Rafols.
\newblock {Indicators of the interdisciplinarity of journals: Diversity,
  centrality, and citations}.
\newblock {\em Journal of Informetrics}, 5(1):87--100, 2011.

\bibitem{Liu:2014}
Yu~Liu, Zhengwei Sui, Chaogui Kang, and Yong Gao.
\newblock Uncovering patterns of inter-urban trip and spatial interaction from
  social media check-in data.
\newblock {\em PLoS ONE}, 9(1):e86026, Jan 2014.

\bibitem{Morillo:2003}
Fernanda Morillo, Mar\'{\i}a Bordons, and Isabel G\'{o}mez.
\newblock Interdisciplinarity in science: A tentative typology of disciplines
  and research areas.
\newblock {\em J. Am. Soc. Inf. Sci. Technol.}, 54(13):1237--1249, November
  2003.

\bibitem{Pedersen:2016}
David~Budtz Pedersen.
\newblock Integrating social sciences and humanities in interdisciplinary
  research.
\newblock {\em Palgrave Communications}, 2:16036, 2016.

\bibitem{Rons:2013}
Nadine Rons.
\newblock Interdisciplinary research collaborations: Evaluation of a funding
  program.
\newblock {\em Collnet Journal of Scientometrics and Information Management},
  5, 07 2013.

\bibitem{Sayama:2012}
Hiroki Sayama and Jin Akaishi.
\newblock Characterizing interdisciplinarity of researchers and research topics
  using web search engines.
\newblock {\em PLOS ONE}, 7(6):1--9, 06 2012.

\bibitem{Chen:2015}
Chen Shiji, Gingras Yves, Arsenault Clément, and Larivière Vincent.
\newblock Interdisciplinarity patterns of highly‐cited papers: A
  cross‐disciplinary analysis.
\newblock {\em Proceedings of the American Society for Information Science and
  Technology}, 51(1):1--4, 2015.

\bibitem{Singh:2017}
Mayank Singh, Rajdeep Sarkar, Pawan Goyal, Animesh Mukherjee, and Soumen
  Chakrabarti.
\newblock Relay-linking models for prominence and obsolescence in evolving
  networks.
\newblock In {\em Proceedings of the 23rd ACM SIGKDD International Conference
  on Knowledge Discovery and Data Mining}, KDD '17, pages 1077--1086. ACM,
  2017.

\bibitem{Solomon:2016}
Gregg E.~A. Solomon, Stephen Carley, and Alan~L. Porter.
\newblock How multidisciplinary are the multidisciplinary journals science and
  nature?
\newblock {\em PLOS ONE}, 11(4):1--12, 04 2016.

\bibitem{Lariviere:2010}
Lariviere Vincent and Gingras Yves.
\newblock On the relationship between interdisciplinarity and scientific
  impact.
\newblock {\em Journal of the American Society for Information Science and
  Technology}, 61(1):126--131, 2010.

\bibitem{Weart:2012}
Spencer Weart.
\newblock Rise of interdisciplinary research on climate.
\newblock {\em Proceedings of the National Academy of Sciences}, 110(Supplement
  1):3657--3664, 2013.

\bibitem{xu:2011}
Yuesheng Xu, Haizhang Zhang, and Qinghui Zhang.
\newblock Refinement of operator-valued reproducing kernels, 2011.

\bibitem{Yegros-Yegros:2015}
Alfredo Yegros-Yegros, Ismael Rafols, and Pablo D’Este.
\newblock Does interdisciplinary research lead to higher citation impact? the
  different effect of proximal and distal interdisciplinarity.
\newblock {\em PLOS ONE}, 10(8):1--21, 08 2015.

\end{thebibliography}
\end{document}